# A Novel Deep Learning Based Approach for Left Ventricle Segmentation in Echocardiography: MFP-Unet

Shakiba Moradi[1,2#], Mostafa Ghelich Oghli[1,3#*], Azin Alizadehasl[4], Isaac Shiri[5], Niki Oveisi[6], Mehrdad Oveisi[3,7], Majid Maleki[3], Jan Dhooge[8]

1- Research and Development Department, Med Fanavarn Plus Co., Karaj, Iran

2- Department of Electrical Engineering, Sharif University of Technology, Tehran, Iran.

3- Rajaie Cardiovascular Medical and Research Center, Iran University of Medical Science, Tehran, Iran.

4- Echocardiography and Cardiogenetic research centers, Cardio-Oncology Department, Rajaie Cardiovascular Medical & Research Center, Tehran, Iran

5- Division of Nuclear Medicine and Molecular Imaging, Geneva University Hospital, CH-1211 Geneva 4, Switzerland

6- School of Population and Public Health, The University of British Columbia, BC, V6T 1Z4, Canada.

7- Department of Computer Science, University of British Columbia, Vancouver, BC, V6T 1Z4, Canada.

8- Department of Cardiovascular Sciences, KU Leuven, Leuven, Belgium.

\# Both authors contributed equally to this manuscript.

*Corresponding author: Mostafa Ghelich-Oghli
Postal address: 10th St. Shahid Babaee Blvd. Payam Special Economic Zone, Karaj, Iran.
Postal code: 3187411213
Tel: (+98) 26- 34239835-41
Fax: (+98) 26- 34239842 (110)
Email: m.g31.mesu@gmail.com

# ABSTRACT

Segmentation of the Left ventricle (LV) is a crucial step for quantitative measurements such as area, volume, and ejection fraction. However, the automatic LV segmentation in 2D echocardiographic images is a challenging task due to ill-defined borders, and operator dependence issues (insufficient reproducibility). U-net, which is a well-known architecture in medical image segmentation, addressed this problem through an encoder-decoder path. Despite outstanding overall performance, U-net ignores the contribution of all semantic strengths in the segmentation procedure. In the present study, we have proposed a novel architecture to tackle this drawback. Feature maps in all levels of the decoder path of U-net are concatenated, their depths are equalized, and up-sampled to a fixed dimension. This stack of feature maps would be the input of the semantic segmentation layer. The proposed network yielded state-of-the-art results when comparing with results from U-net, dilated U-net, and deeplabv3, using the same dataset. An average Dice Metric (DM) of 0.945, Hausdorff Distance (HD) of 1.62, Jaccard Coefficient (JC) of 0.97, and Mean Absolute Distance (MAD) of 1.32 are achieved. The correlation graph, bland-altman analysis, and box plot showed a great agreement between automatic and manually calculated volume, area, and length.

# INTRODUCTION

Cardiovascular diseases (CVDs) are the main cause of mortality around the world [1]. The first step for the control and treatment of these diseases is an accurate diagnosis, which is achievable by diagnostic tools such as imaging systems (e.g. echocardiography, angiography, magnetic resonance imaging, etc.). Among these imaging systems, echocardiography undoubtedly is the preferred device for the evaluation of cardiac ventricles. Although cardiac magnetic resonance imaging (CMRI) provides a better visualization of anatomical structures and wall motion, echocardiography is more popular due to its low cost, temporal resolution, and portability [2]. To have an efficient diagnosis, there is a crucial need for left ventricle (LV) segmentation to be used in calculations of clinical indices such as end diastolic and end systolic volumes, ejection fraction, left ventricular mass, etc. [3]. Routinely, an expert cardiologist manually delineates the LV endocardial border at end diastole (ED) and end systole (ES) frames of a cardiac cycle. This procedure is not only a tedious and time-consuming task, but also suffers from poor reproducibility and intra and inter-observer variability [4]. To address this, an automatic segmentation method is required to accelerate the process while reducing inter-user variability. However, LV segmentation problem faces with several challenges that a robust, reliable method could overcome. Firstly, some of the characteristics of ultrasound (US) images (e.g. low signal to noise ratio, weak echoes, the presence of speckle, etc.) produces images with unclear borders, which render the delineation of LV extremely difficult in some cases. Second, the determination of the region of interest (i.e. the LV region) is challenging due to the presence of more than one anatomical structure in echocardiographic images. Finally, there is no simple relation between the physical property of the tissue and the pixel intensity in images.

There are several methods which address the ventricle segmentation problem, including active contour, active shape and appearance methods, bottom-up approaches, and machine learning based methods [3–11]. Most of these methods focus on endocardial border detection in a single frame echocardiography image. An early survey of LV segmentation methods are presented in [3]. Active contour approaches were popular for this purpose in the early 2000s [7,12,13]. Yan et al. [12] used a modified fast marching method for solving active contour problem in echocardiography image segmentation. They utilized an average intensity gradient-based measure within the speed term of a fast-marching model. This was done to protect the contour energy conception from noise disruption and increase the speed of advancement to the object boundary. Since the proposed method is applied to short axis images, the main concern about this method is the capability of delineation of the LV in apical 4-chamber view.

To improve segmentation accuracy, the regional, edge, and shape information is incorporated into the active contour equation in numerous works. For instance, region and edge information of echocardiography images are utilized in the active contour equation in [7]. In another work, Chen et al. [13] constrain evolving active contours to a shape prior term to find boundaries that are similar to the prior, even when the whole boundary is unclear in imaging. Although utilizing this approach is useful in echocardiographic images, the paper suffers from a lack of comprehensive datasets to create the shape prior.

The active appearance model (AAM) was another frequently used approach for the LV border delineation in echocardiography images. Bosch et al. [5] introduced an extended version of the AAM that they called the active appearance motion model (AAMM) to delineate the LV endocardial contours over the full heart cycle. The intensity of the echocardiographic sequences was normalized nonlinearly to augment the advantages of AAM with temporal consistency. Furthermore, Mitchell et al. [6] developed a 3D AAM (2D+time) to segment LV in cardiac MR and echocardiographic images. Their work requires no additional interactively supplied information and the training phase was implemented by a series of manually segmented examples. The main drawback of AAM is using a point distribution model (PDM) for the shape representation. This is because PDM requires point correspondence which can be very tedious and challenging in medical data. Moreover, AAM and active contour perform effectively only when prior knowledge about the LV shape and appearance are present in the method [14].

The advent of deep neural networks [15,16] affects all aspects of computer vision and accordingly medical image analysis, such as image to image translation [17], noise reduction [18], image reconstruction [19] and super resolution image generation [20, 21]. Carneiro et al. [4] proposed a deep learning approach based on deep belief network (DBN) that decouples the rigid and nonrigid classifiers for segmentation of LV in 4-chamber view echocardiographic images. The results of training and evaluation of their network on 480 annotated images from 14 sequences show promise. Smistad et al. [22] pre-trained U-net [23] using an automatic Kalman filter (KF) and compared the results with the KF method. DM [24] and HD [25] are used for the evaluation of the results regarding manual delineation. Due to this, the DMs of U-net and KF were quite similar and the HD of the proposed method performed better significantly. Zyuzin et al [26] utilized U-net architecture for the segmentation of LV in 2D echocardiography images in 4-Chamber view. They used DM to evaluate their work and achieved a DM of 0.923 for augmented data. Their data contained 188 frames of 4-chamber view images and was augmented by shifting and scaling methods.

Among the deep learning-based methods, the convolutional neural network (CNN) is able to handle US image segmentation challenges to a high extent. This is due to its shift invariance characteristics and strong capability in image feature extraction. In this paper, we propose a novel approach based on CNN in 2D echocardiographic images for the delineation of LV endocardium border. U-net, which showed promising results in medical image segmentation for various applications [27–30], is built and modified upon the fully convolutional network [31]. It proposes two main innovations: network symmetry and skip connections between the encoder-decoder paths. These aspects concatenate encoding feature maps to the corresponding decoding feature maps. Despite ingenuity, U-net ignores the effect of feature maps in different scales "directly". Due to this, when the feature maps are going to up-sample in the decoder path, only the last layer connects to the output segmentation map, as any other feedforward network. Additionally, blocks in different scales of the decoder path cannot share features, which may prevent the flow of information and result in superfluous parameters. While it's possible to provide the input of the last layer using all steps of the decoder path.

This study proposes using feature pyramid networks (FPN) [32] to address this issue. Feature pyramids are a basic component in recognition systems for detecting objects at different scales. The key idea is extracting feature maps from all the levels of the decoder path separately and use them for final pixel classification. The feature pyramids in our proposed model are similar to FPN, but differ in their design philosophy from feature extractor to semantic levels.

The proposed method is compared with three segmentation approaches: (i) U-net, (ii) dilated U-net [33], and (iii) deeplabv3 [34]. Deeplab [34–36] is a novel approach for semantic segmentation that is based on three main innovations:

1- Atrous/dilated convolution: convolution with upsampled filters.

2- Atrous spatial pyramid pooling (ASPP): filters convolution feature layer with multiple sampling rates.

3- Fully connected conditional random field (CRF): a probabilistic graphical model that improves the localization of object contour.

The reason for comparing achieved results with the deeplab network is the usage of ASPP in both networks. Based on kaggle leaderboard, deeplabv3 has yielded the best results for segmentation in the VOC2012 dataset.

In the following sections the base networks containing U-net, dilate U-net, and FPN are described, the proposed architecture is defined, the dataset is introduced, and the results are presented.

## ARCHITECTURE OF BASE NETWORKS: U-NET AND FPN

The U-net architecture consists of an encoder path and a decoder path, as illustrated in Fig. 1. The network is composed of convolution operation, max pooling, ReLU activation, concatenation, and up convolution layers. Initially, the input image is fed into the network. The data is then propagated through the network along all possible paths (contraction, expansion, and concatenation). At the end, the binary segmentation will appear. Each orange box in Fig. 1 corresponds to a multi-channel feature map and the spatial size is denoted at the bottom of each box. Adjacent boxes represent convolution operation followed by a nonlinear activation

function— i.e. rectified linear unit (ReLU). Each convolution layer contains a set of parameters which form learnable filters (also known as kernels). These filters have a small receptive field with a depth equal to the depth of the input volume. The convolution process is multiplying the entries of the filter with the input image pixel values which means element-wise multiplications. This procedure produces a 2-dimensional feature map of that filter. The formulation of convolution layer is:

$$x_{ij}^{l} = \sum_{a=0}^{m-1}\sum_{b=0}^{m-1} w_{ab} y_{(i+a)(j+b)}^{l-1} \qquad (1)$$

where the filter size is m×m, the weights are $w_{ab}$, the output of previous layer is $y^{l-1}$, and the output of this layer is $x^l$.

ReLU activation function applies a simple non-saturating activation function ReLU(x) = max (0, x). Thus, the output of $l^{th}$ layer would be:

$$y_{ij}^{l} = \text{Re}LU(x_{ij}^{l}) \qquad (2)$$

The next operation is max pooling which is a form of non-linear down-sampling that reduces the x-y size of a feature map. The procedure is partitioning input images into a set of non-overlapping regions and determination of maximum for each such sub-region. After each max pooling operation, the number of feature channels is increased by a factor of two. The sequence of

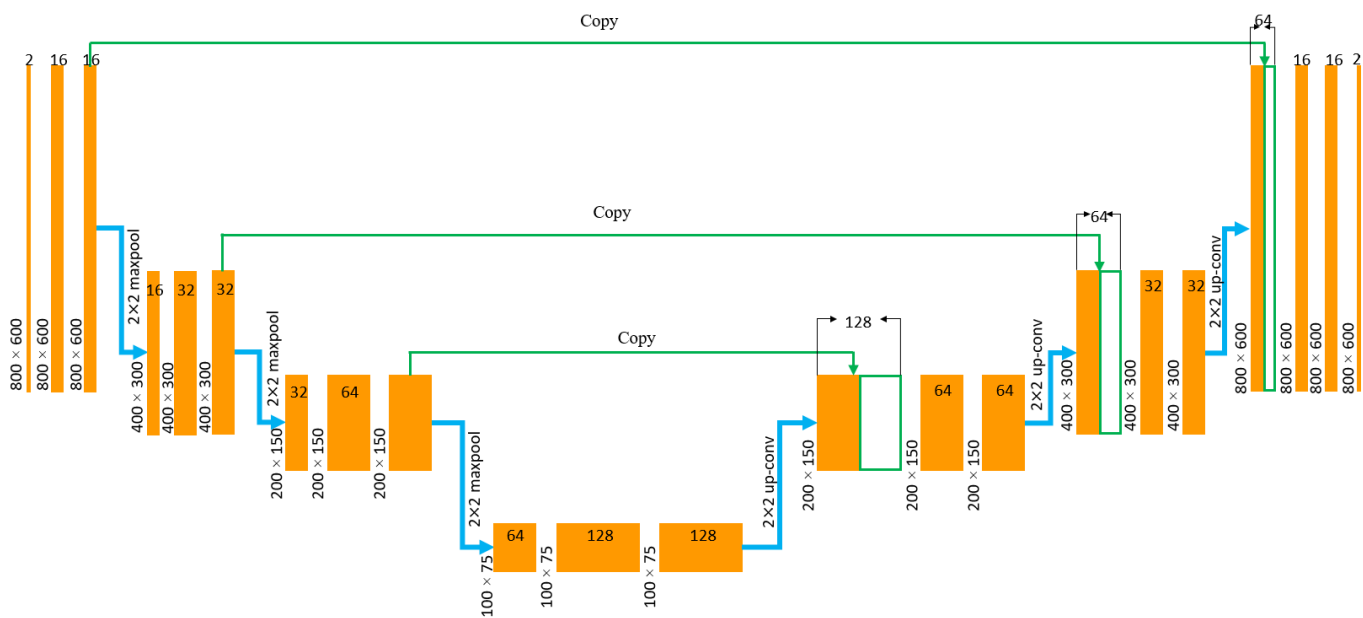

All convolutions are 3×3, Relu

Fig. 1. U-net architecture

convolution and max-pooling operations result in a contraction path that increases the number of feature maps that would be learned. In addition to this, the resolution of feature maps is also decreased. Standard classification networks end here and map all feature vectors to a single output vector. The U-net has an additional "expansion" path to create a high-resolution segmentation map. This expansion path consists of a sequence of up-convolutions and concatenation with corresponding high-resolution features from the contracting path.

The up convolution (or transposed convolution), uses a learned kernel to map each feature vector to the 2×2 output window, again followed by a ReLU activation function. There are also some concatenation paths in the network as seen in Fig. 1. Higher resolution features are added from the contraction path to the corresponding expansion level in order to better localize and learn representations with following convolutions. This concatenation path provides a good prior from earlier stages to cover the sparsities of up-sampling operations.

The output segmentation map has two channels: one for foreground and one for background classes.

*Dilated U-nets: Global Receptive Fields*

Medical image segmentation requires global information of the image to see how organs are arranged relative to one another. U-net could not perceive the entire image in the layers of contraction and expansion paths. This occurs, as the network would have little understanding that there is only one left ventricle in an ultrasound image. A readily available approach to encompass this weakness is adding two more downsampling layers. This method, however, increases network parameters. Due to this, dilated convolution kernels [33] are used in this work to increase the receptive fields of the network. Dilated convolutions space out the pixels summed over in the convolution by a dilation factor.

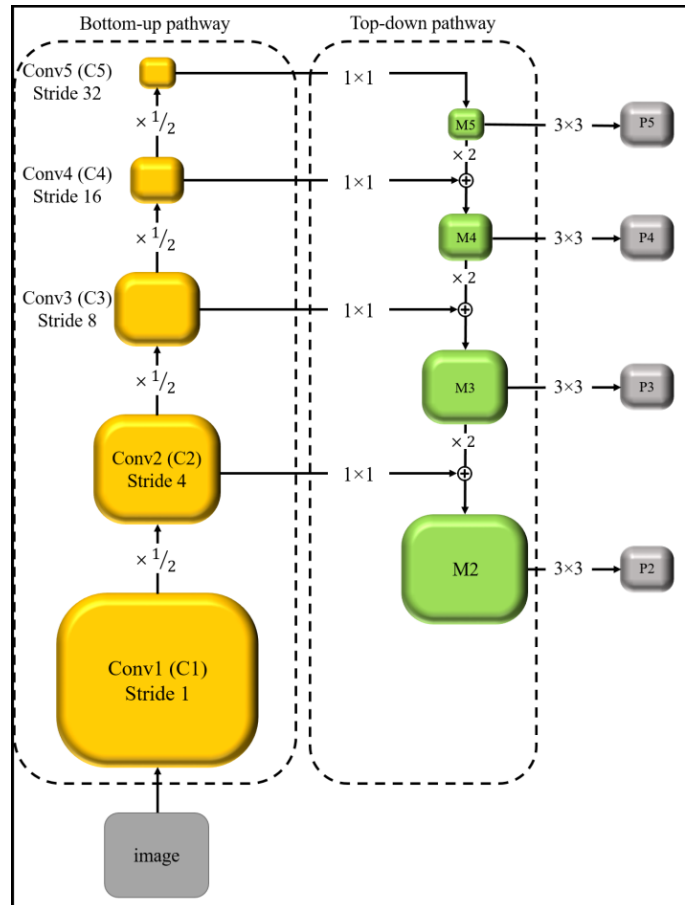

Fig. 2.  Feature pyramid network, representing bottom-up and top-down pathways.

*Feature Pyramid Network*

FPN is an efficient architecture for detecting objects in different scales [32]. There are three approaches to recognizing objects at different scales:

1- Pyramids of the same image at various scales is the first approach [37].
2- Pyramid of features: Regarding the definition of CNNs, the features of images are extracted in them hierarchically. This is the key point that leads the researchers to create a pyramid of features, instead of a pyramid of images.
3- Pyramid of features with the same semantic strength at all scales: This goal is easily achievable by creating an extra path that adds high-level features from the top layers to the bottom layers that have higher resolution, but semantically weak features.

This proposed solution has one weakness which is inaccuracy of the objects location after all the down and up-sampling. This inaccuracy is controlled through adding lateral connections between reconstructed layers and corresponding feature maps. This idea has been popular in recent researches [38–40], where they tried to predict from the highest resolution reconstructed layer. Conversely, FPN utilizes all the reconstructed layers (at all levels) for prediction independently. This key concept inspired this works proposition of a U-net architecture in which feature maps in all of the decoder path levels are incorporated in the segmentation procedure independently, rather than using the topmost level.

As portrayed in Fig. 2, there are two pathways that the input image propagates through. The first one is the "bottom-up pathway" and the second is the "top-down pathway". To define a pyramid of feature maps, the spatial sizes should be reduced regularly from bottom to top to create different levels of the pyramid. It includes several convolution modules named {C1, C2, C3, C4, and C5}. As the input image goes through the network, the spatial dimension is reduced by a factor of 0.5, which is equivalent to doubling the stride. Specifically, there are strides {2, 4, 8, 16, and 32} for corresponding {C1, …, C5}. The C1 module has not been included in the pyramid due to its large memory footprint.

The top-down pathway provides the opportunity of transferring semantically stronger features to a higher resolution dimension. These features are then enhanced with corresponding features of the bottom-up pathway via skip connections. There are some specific processes available to make the bottom-up feature maps appropriate for merging with top-down feature maps. A 1×1 convolution filter is applied to reduce C5 channel depth. Then, a 3×3 convolution is applied to create P5 which is the first feature map used for further prediction. At each level, the newly created feature map is up-sampled and summed with the result of the application of 1×1 convolution to the corresponding feature map in the bottom-up pathway. This procedure is repeated until the last resolution map is created with the highest quality. Fig. 2 demonstrates the process rigorously.

# MFP-Unet: The Proposed Architecture

This study proposes a hybrid architecture for 2-D echocardiographic image segmentation. This architecture is the multi-feature pyramid U-net (MFP-Unet) and it is a combination of dilated U-net and Feature Pyramid Networks. As mentioned before, the contraction path of U-net architecture creates incremental numbers of feature maps, but consequently, resolution will diminish. This is the reason for the application of the concatenation process in each level of expansion path. The combination of deep features' information and low-level features with high resolution allow the algorithm to save both accuracy and resolution. Despite having this architecture, U-net uses only the last layer features in the expansion path for the segmentation step (the last layer). This study proposes the MFP-Unet to modify the U-net architecture so that the feature maps in all levels of expansion path are incorporated separately.

Considering U-net architecture, MFP-Unet adds extra convolution layers for extracting feature maps from all levels of the expansion path in order to be included in the segmentation process in the last layer. This inclusion is promised by a feature pyramid network. In this way, a simple separate path is added to each level of the decoder which is a 3×3 convolution layer with 16 feature maps and an up-sampling layer to make the size of the feature maps same as the last layer. Finally, all the extracted feature maps are concatenated to make a final feature map with 64 channels, and this will be the one shown to the 1×1 convolution filter for the classification process. Fig. 3 shows this architecture in detail.

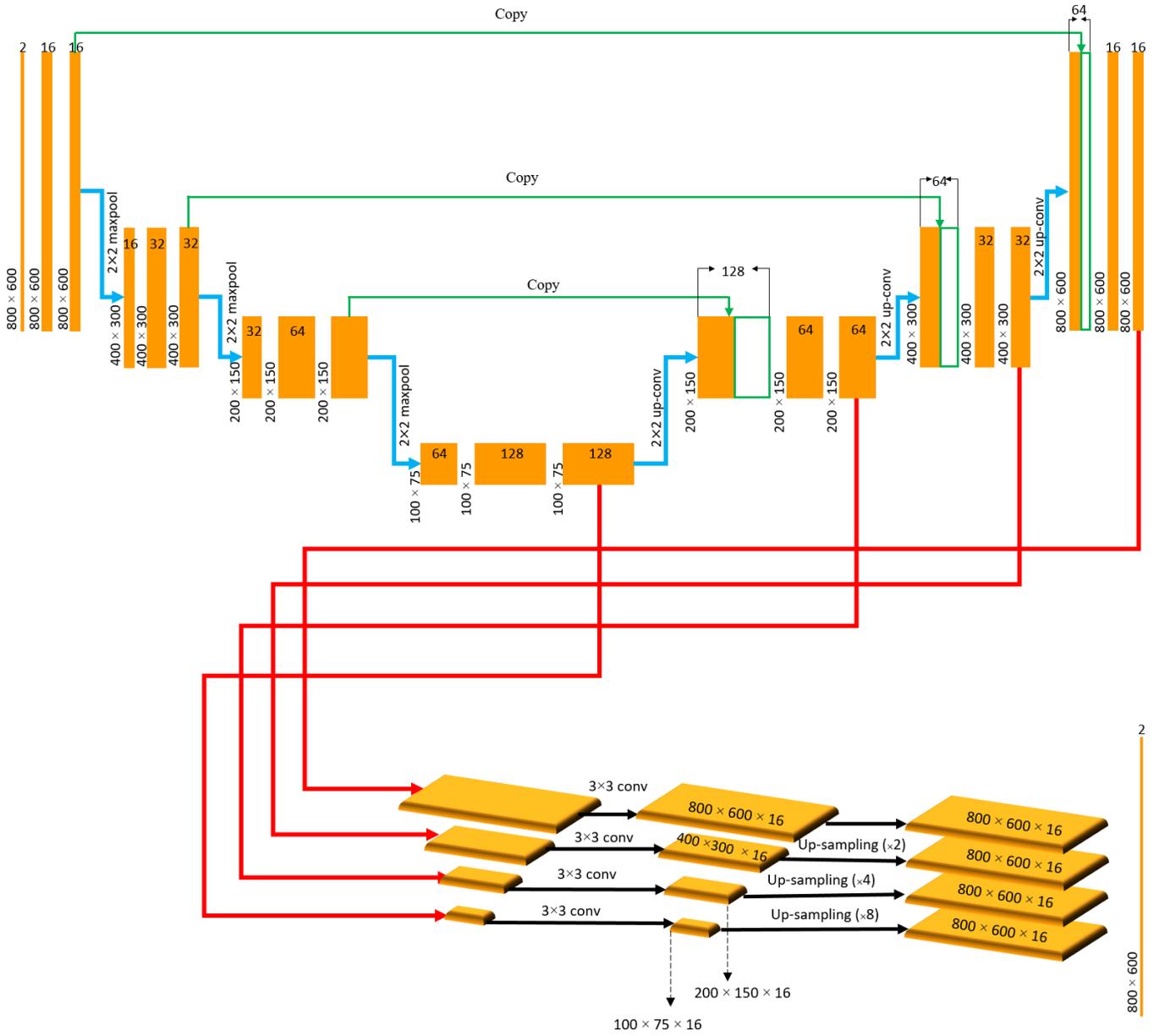

Fig. 3. The proposed MFP-Unet architecture. The expansion path acts as bottom-up pathway of FPN (but vice versa). Therefore, feature maps of all scales are gathered, their depth and dimension are equalized, and finally the segmentation step is performed.

To formulate the procedure, the last layer in each level of the decoder path is named up$^i$ (i = 1, 2, 3, 4). First, a closer look at the last layer of the classic U-net (i.e. fully connected layer) would be beneficial. As the fully connected layer is a convolution layer with a 1×1 kernel, its output would be calculated regarding equations 1 and 2 with m=1 and l=L (the last layer).

$$y_{ij}^L = \operatorname{Re} LU(w_{00} y_{ij}^{L-1}) \qquad (3)$$

According to the classic U-net architecture, the represents the up$^4$. However, the equation differs in the MFP-Unet:

$$y_{ij}^{L-1} = concat \begin{bmatrix} conv(up^4,16), conv(up^3,16) \\ conv(up^2,16), conv(up^1,16) \end{bmatrix} \quad (4)$$

Where the *concat[…]* is the concatenation operation and the *conv(up$^i$,16)* is the convolution unit (convolution and ReLU layers) performed on the *up$^i$* with 16 feature maps in output.

The encoder path remains unchanged, but dilated convolution filters were used to be more robust in dilated U-net. In the decoder path, the layers of U-net architecture remained unchanged, but extra layers are added to improve the accuracy and resolution of each level in separate feature maps. Finally, instead of solving the segmentation problem by having just the last layer feature map, the saved feature maps were combined and shown to the network classifier.

Although the U-net architecture is simply modified in our work, it will overcome the resolution and accuracy problem in the encoder-decoder paths. In other words, U-net contraction and expansion paths act as the FPN bottom-up and top-down pathways, the decoder feature maps are gathered directly, and after applying convolution layers and up-sampling the final layer (classifying each pixel as foreground or background) is performed.

*Input extra channel*

Image preprocessing is an important step in decision-making tasks and computer vision. This procedure may include operations for noise cancelation, contrast enhancement, etc. All these operations have a variety of techniques which are used according to the image type and purpose of the preprocessing. This study is working with grayscale medical images of echocardiography in 4 chamber view and the purpose is to delineate LV borders. The LV borders are the lateral wall and septum which have high acoustic impedances so they will appear brighter in ultrasound imaging technique. Due to this, a preprocessing method for contrast enhancement to make a good margin between the borders and other bright pixels in the image is used in this work.

One way to enhance the contrast in images is thresholding. It is an effective and simple tool in many applicants where the gray levels of objects in the image are different from the ones belonging to the background pixels. Thresholding is performed in two main routines: global and local. In global thresholding, there is one threshold for the whole image while in local thresholding, all pixels are compared to a threshold which is calculated based on some statistics of the neighborhood pixels within a window [41].

One of the popular techniques in local thresholding is the Niblack's technique which is firstly introduced in [42]. In this method, the value of the threshold is determined using mean and standard deviation of the pixels within a window, according to the following equation:

$$T(x,y) = m(x,y) + k \times \delta(x,y) \qquad (5)$$

Where, T(x,y) is the value of the threshold, m(x,y) and δ(x,y) are the local mean and standard deviation, and k is a bias parameter that can be chosen according to the image type.

Here we have used this technique for global thresholding to make the borders brighter in echocardiography images. The values of mean and standard deviation are calculated considering the whole image and k is set 2. The new image which is obtained from passing the original one through the thresholding filter is used as a second channel of inputs to the neural network.

*Left Ventricle Parameter Measurements*

Left ventricle segmentation is an essential tool for studying the left heart function. In this section, the calculation of left ventricle functional parameters will be explained, which are: length, area, volume, and ejection fraction (EF) from the segmented area in 4 chamber view images. The LV length delineation procedure will be described first as it is a prerequisite for the calculation of LV area and volume.

1- The LV contour is delineated by the proposed method.
2- The smallest peripheral triangle, which contains all the border points, is found (cyan triangle in Fig. 4).
3- The nearest point of the border to the three vertices of the triangle is found. These three points determine the mitral valve annulus and the apex.
4- A line that crosses two points around the LV annulus (the blue points in Fig. 4) is detected.
5- The line perpendicular to the baseline is depicted (the blue line between two red circles).
6- The intersection point between the left ventricle contour and the perpendicular line is found.
7- The LV length is the distance between the head of the left ventricle and the intersection point.

The left ventricle area can be calculated by counting the number of pixels in the segmented area multiply by the pixel dimensions.

Calculating the left ventricle volume needs to have a 3D image or multiple image planes from different views. However, if there is no abnormality in left ventricle wall motions, it is possible to estimate the volume using just one view image [2]. As all of the patients in this study have normal left ventricles, this simplification can be used to measure the volume as in the equation below. S and D are the measured area and length, respectively.

$$V = \frac{8S^2}{3\pi D} \tag{6}$$

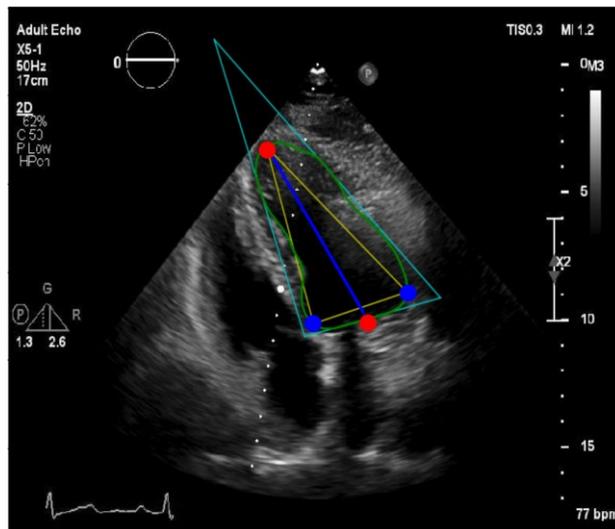

Fig. 4. The LV length measurement procedure. The length is defined as the blue line between the above red circle and the triangle's baseline.

The left ventricle volume is calculated in the end diastolic and end systolic phases of cardiac cycle. Then, the EF is calculated as the difference of end diastolic and end systolic volumes, divided by end diastolic volume.

*Evaluation metrics*

This section presents a statistical evaluation of the proposed LV endocardium segmentation algorithm. The proposed model is also compared with U-net, dilated U-net, and deeplabv3. This study utilized four performance metrics: Dice metric (DM) [24], Hausdorff distance (HD) [25], Jaccard coefficient (JC) [43], and mean absolute distance (MAD) [44] between manually drawn and predicted contours. The DM measures the contour overlap given areas of manually segmented contour ($A_M$), automatically segmented contour ($A_A$), and their intersection.

$$DM = \frac{2(A_A \cap A_M)}{A_A + A_M} \quad (7)$$

This formulation gives a number which is always between zero and one. Lower DM indicates that there is a low intersection between automatic segmentation and the reference ground truth, which means poor segmentation. On the contrary, higher DM denotes a more complete match between automatic segmentations and ground truth. HD measures the maximum perpendicular distance between the automatic and manual contours. A and B denote the two contours. The HD is defined as:

$$HD = \max\left(\max_{a \in A}\left(\min_{b \in B} d(a,b)\right), \max_{b \in B}\left(\min_{a \in A} d(a,b)\right)\right) \quad (8)$$

where d(.,.) represents Euclidean distance. JC is similar to DM and is used for comparing the similarity and diversity of segmented areas. The JC measures the similarity between two segmentations, and is defined as the number of pixels of the intersection area divided by the number of pixels of the union area:

$$(9)$$

The MAD uses the same distance metric, d(a, b), as defined in HD between two given contours

A, B. MAD is calculated as the average distance from all points on the contour of the automatically segmented LV to the contour of the ground truth [44].

$$MAD = \frac{1}{n_a} \sum_{i=1}^{n_a} |d(a,b)| \quad (10)$$

Where $n_a$ is the number of points on automatic contour.

# RESULTS

*Prepared dataset for training and evaluation and manual annotation protocol*

A collection of 137 2D echocardiographic transthoracic four-chamber sequences were prepared. To ensure that the training procedure is performed with a satisfactory variety of images, different pathological conditions are also included to the image dataset in addition to the normal cases (e.g. pulmonary thromboembolism, pulmonary hypertension, cardiomyopathies, etc.). As the dataset is not large enough to be appropriate for training the network, the elastic deformation method is used to augment the data by a factor of 10. The images were gathered in the Rajaie Cardiovascular Medical and Research Center and Intelligent Imaging Technology Research Center. The ultrasound machines are EPIQ 7 echocardiographic system (Philips, Amsterdam, Netherlands) with an X5-1 xMatrix transducer (Philips, Amsterdam, Netherlands) and Luna echocardiographic system (Med Fanavaran Plus, Karaj, Iran) with a P2-4 transducer (Vermon, Tours, France). The X5-1 transducer consists of 3040 piezoelectric elements and permits real-time 2D, M-Mode, and 3D image acquisition and rendering. Patients with arrhythmia were not included. Two observers delineated the endocardium boundary manually at ED and ES phases of cardiac cycle using 3Dslicer [45]. Volumes and EF were computed from manual segmentation using a spline-based geometry model [46] and used as a manual reference. The total dataset was split randomly into a training and evaluation set of 1080 images and a testing set of 290 images (through consideration of data augmentation). The calibration factors varied from one data set to another, but in general, with 0.23 to 0.36 mm/pixel and images were captured at a resolution of 800×600 pixels. The research participants were notified about being as a part of this study and the study was approved by the ethics committee of the research lab.

*Implementation details*

A total of 1370 slices with LV were used, including 1080 slices for training and 290 slices for validation. To evaluate and generalize the performance of the model to an independent dataset, the unbiased 5-fold cross-validation method is considered and applied at the image level. For this purpose, echocardiographic images in the study are first stratified and partitioned into 5 equally sized folds to ensure that each fold is a good representative of the whole. Subsequently, 5 iterations of training and validation are performed such that within each iteration a different fold of the data is held out for validation while the remaining 4 folds are used for the learning process.

The MFP-Unet is trained with Stochastic Gradient Descent (SGD) with a momentum of 0.9 and a weight decay of 0.0005. Training parameters in the learning phase are: learning rate = 0.001, batch size = 64, max epoch = 100, and decay = 10e−5).

The processing hardwares were 12GB of RAM, a GPU based graphic card with 2496 CUDA cores (Tesla K80), and an Intel Xeon CPU. The network implementation was done in the Python environment with Tensorflow r1.12 and Keras 2.2.4.

*Manual and automatic LV Segmentation Reproducibility*

The reproducibility analysis of the manual and automatic segmentation is provided in Tables 1 and 2. Two expert radiologists blindly performed manual segmentations of LV endocardial contours to derive inter-observer variability. To evaluate the inter-observer variability of the proposed model, segmentation results are compared against the mean of the users' segmentation output. All inter-observer variability results were statistically significant in manual fashion and are considerably reduced in the automatic fashion. This can be perceived by the reduction of p-values for all the clinical parameters.

The Bland-Altman plots of volume, area, length, and EF are also illustrated in Figs. 6-9 for assessing the intra and inter-observer variability. The reproducibility coefficient (RPC) and the coefficient of variation (CV) are also computed from the Bland-Altman analysis to quantitate the intra and inter-observer variability. The RPC is defined as 1.96∗SD, and the CV is defined as:

$$CV = \frac{SD(auto - man)}{mean(auto) + mean(man)} \quad (11)$$

Where *SD (auto – man)* is the standard deviation (SD) of the differences between the automatic and manual results, while mean (*auto*) and mean (*man*) are their mean values.

The mean and confidence intervals of the difference between the automatic and proposed method volume results were -2.46 mL and -24.28 mL to 19.35 mL, respectively. CV and RPC values were 5.71% and 21.81%, respectively. The mean and confidence interval of the difference between the automatic and proposed method area results were -0.3 cm² and -4.02 cm² to 3.4 cm², respectively. CV and RPC values were 3.19% and 3.71%, respectively. The mean and confidence interval of the difference between the automatic and proposed method length results were 0.01 cm and -0.63 cm to 0.66 cm, respectively. CV and RPC were 2.1% and 0.65%, respectively. The mean and confidence interval of the difference between the automatic and proposed method EF results were -0.88% and -14.79 mL to 13.01 mL, respectively. CV and RPC were 10.71% and 13.9% respectively.

TABLE I
Manual left ventricle segmentation reproducibility. The first column shows mean values and the range in parenthesis. The second column shows the corresponding percentages. Finally, the last column demonstrates equivalent p-values.

|  | Value | Percentage | P |
| --- | --- | --- | --- |
| Volume | 11.62 ml (1.63 to 23.12 ml) | 3.74% (0.79 to 4.75%) | 0.0012 |
| Area | − 4.73 cm² (− 8.32 to 7.45 cm²) | 2.56% (0.49 to 6.05%) | 0.0358 |
| Length | 1.08 cm (0.11 to 2.32 cm) | 9.63% (1.2 to 21.93%) | 0.0024 |
| EF | 18.23% (− 22.86 to 28.92%) | 1.42% (− 1.38 to 2.41%) | 0.0928 |

TABLE II
Automatic left ventricle segmentation reproducibility. The first column shows mean values and the range in parenthesis. The second column shows the corresponding percentages. Finally, the last column demonstrates equivalent p-values.

|  | Value | Percentage | P |
| --- | --- | --- | --- |
| Volume | − 2.46 ml (− 11.02 to 8.87 ml) | 0.67% (− 0.019 to 1.14%) | 0.0009 |
| Area | − 0.3 cm² (− 3.98 to 1.91 cm²) | 0.95% (− 0.11 to 3.28%) | 0.0174 |
| Length | 0.01 cm (− 0.58 to 0.67 cm) | 0.11% (− 5.3 to 6.16%) | 0.0014 |
| EF | − 0.88% (− 13.53 to 5.02%) | 1.01% (− 0.78 to 2.23%) | 0.0073 |

*Illustrative results*

In Table 3, the average values and the standard deviation of the computed metrics are listed for the train and test datasets. Table 4 presents a comparison of results between this method and previously presented methods containing: U-net, dilated U-net, and deeplabv3 using the same database.

An analysis of variance (ANOVA) test is performed on the achieved evaluation results [47] in Table 5. This analysis tests the null hypothesis that the mean value of similarity metrics does not differ significantly. This is compared against the alternative hypothesis that the mean value of similarity metrics differs significantly. In this way, the ANOVA table shows the between-groups and within-groups variation. In Table 5, SS is the sum of squares and df is the degrees of freedom. The total degrees of freedom are total number of observations minus one, which is 16 - 1 = 15.

TABLE III
Evaluation metrics of our proposed method for validation and test sets of our prepared database. Numbers' format: mean ±Sd (standard deviation).

| Dataset | DM[1] | HD[2] | JC[3] | MAD[4] |
|---|---|---|---|---|
| validation | 0.97±0.13 | 1.12±0.23 | 0.95±0.11 | 1.01±0.18 |
| test | 0.945±0.12 | 1.62±0.05 | 0.96±0.01 | 1.32±0.53 |

[1] Dice Metric (DM)
[2] Hausdorff Distance (HD)
[3] Jaccard Coefficient (JC)
[4] Mean Absolute Distance (MAD)

TABLE IV
Comparison of segmentation performance between the proposed method and related techniques using the same database and other methods. Numbers format: mean value ± Sd(standard deviation).

| Method | DM | HD | JC | MAD |
|---|---|---|---|---|
| **Proposed method** | **0.945±0.12** | **1.62±0.05** | **0.98±0.01** | **1.32±0.53** |
| U-net | 0.89±0.21 | 2.20±4.12 | 0.91±0.11 | 3.08±4.63 |
| Dilated U-net | 0.917 ± 0.14 | 1.68 ± 0.05 | 0.93±0.01 | 1.4 ± 0.49 |
| Deeplabv3 | 0.915 ± 0.14 | 1.69 ± 0.05 | 0.92±0.011 | 1.97 ± 0.87 |

TABLE V
Analysis of variance (ANOVA) test of the evaluation metrics.

| Source | SS | df | MS | F | p-value |
|---|---|---|---|---|---|
| between-groups variation | 3.524 | 3 | 1.17 | 6.41 | 0.0077 |
| within-groups variation | 2.198 | 12 | 0.183 | | |
| Total | 5.723 | 15 | | | |

The between-groups degrees of freedom are number of groups minus one, which is 4 - 1 = 3. The within-groups degrees of freedom are total degrees of freedom minus the between groups degrees of freedom, which is 15 - 3 = 12. MS is the mean squared error, which is:

$$V = \frac{SS}{df} \quad (12)$$

for each source of variation. The F-statist ic is the ratio of the mean squared errors. The p-value is the probability that the test statistic can take a value greater than the value of the computed test statistic. The small p-value of 0.0077 indicates that differences between metrics' means are significant.

Fig. 5 shows some of the LV endocardium segmentation results of the MFP-Unet and a comparison with other related methods consist of U-net and deeplabv3. It demonstrates that the segmentation procedure has successfully constrained the LV region better than all the other methods.

After computing four clinical parameters (i.e. length, area, volume, and EF) using the automatic and manual LV segmentation results, they are then used for the correlation and Bland-Altman analyses [48]. The correlation analysis is performed to achieve the slope, intercept, and correlation coefficient R-value.

Figs. 6-9 illustrate the correlation graphs (left) between the automatic and manual results and the Bland-Altman graphs (right) of the differences, using the test dataset for length, area, volume, and EF of the segmented LV, respectively. As the figures show, correlation with the ground truth contours of 0.97, 0.97, 0.91, and 0.85 for volume, area, length, and EF were measured,

respectively.

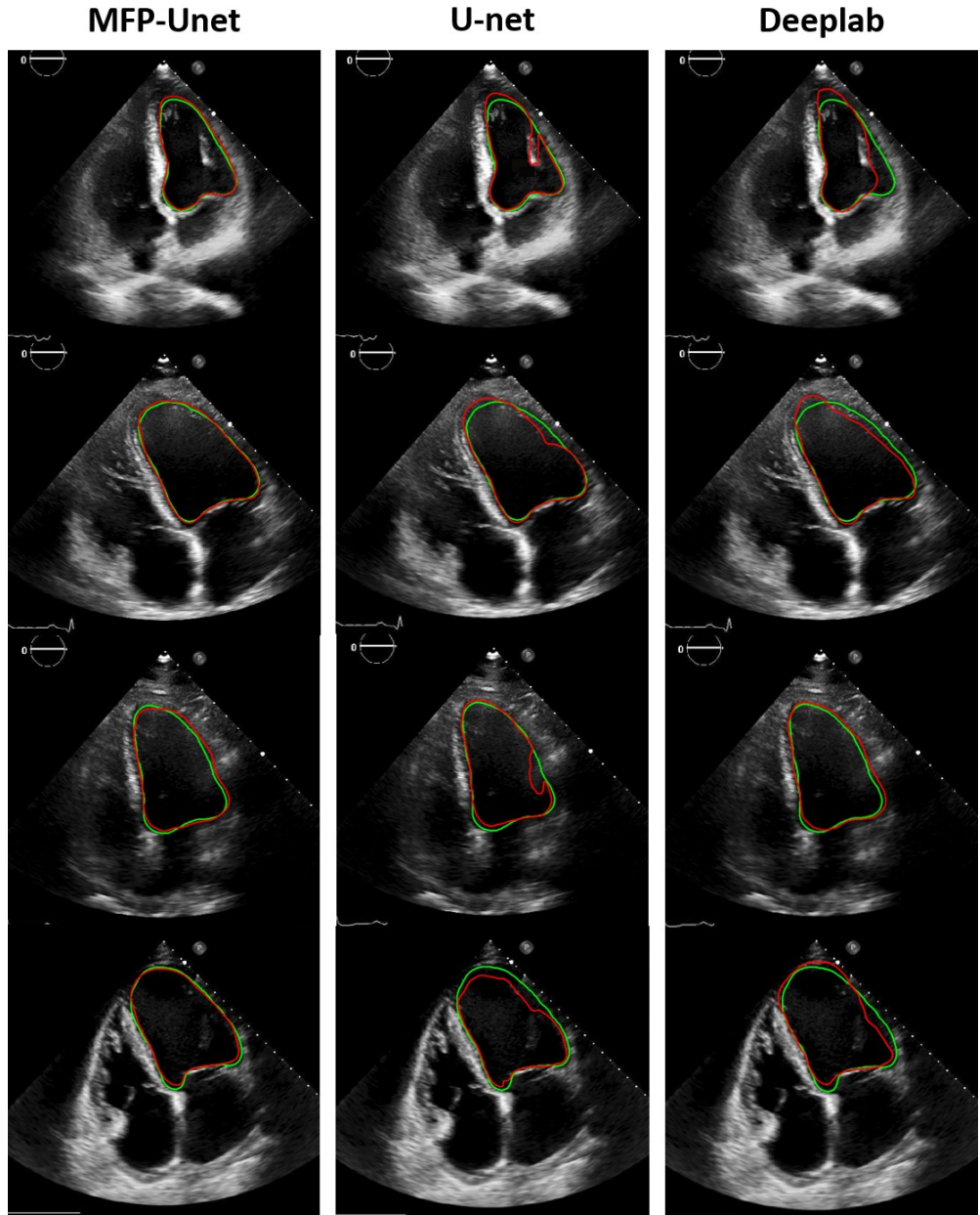

Fig. 5. Some of the sample 4-chamber echocardiographic images and the segmented left ventricle with the proposed method, U-net, and deeplabv3. The green contour relates to ground truth segmentation and the red contour demonstrates the result of automatic segmentation.

Fig. 10 shows the box plots of absolute error of LV functional parameters (volume, area, length, and EF) for each presented approach. The whiskers were plotted by a "plus" sign. The volume and area error of MFP-Unet is the least in comparison with the other approaches. However, its height value is a tad higher, which was expected due to the sensitivity of the algorithm of determining length to triangle position finding.

**Volume**

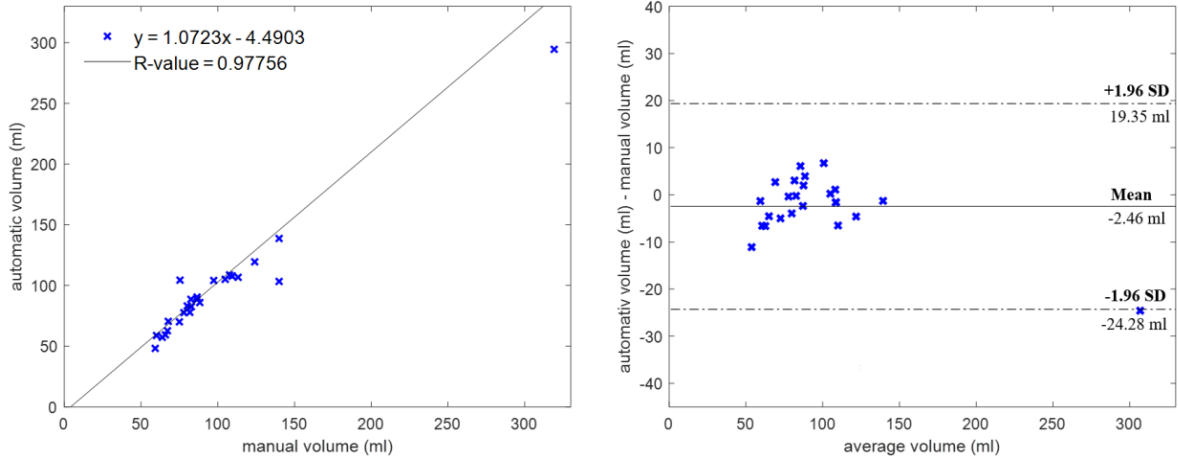

Fig. 6. Correlation graph (left) and Bland-Altman for volume in Test set.

**Area**

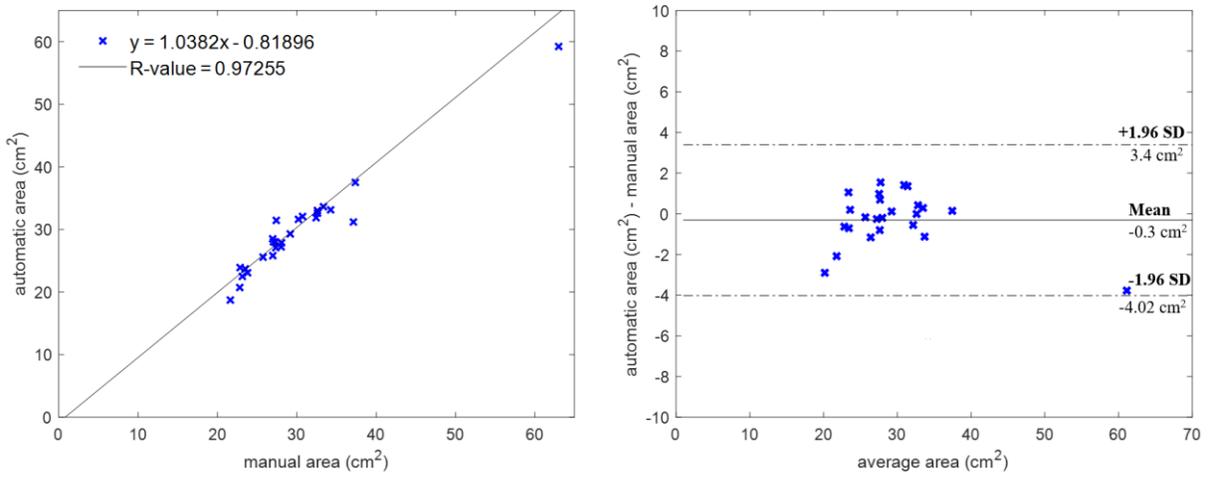

Fig. 7. Correlation graph (left) and Bland-Altman for area in Test set.

**Length**

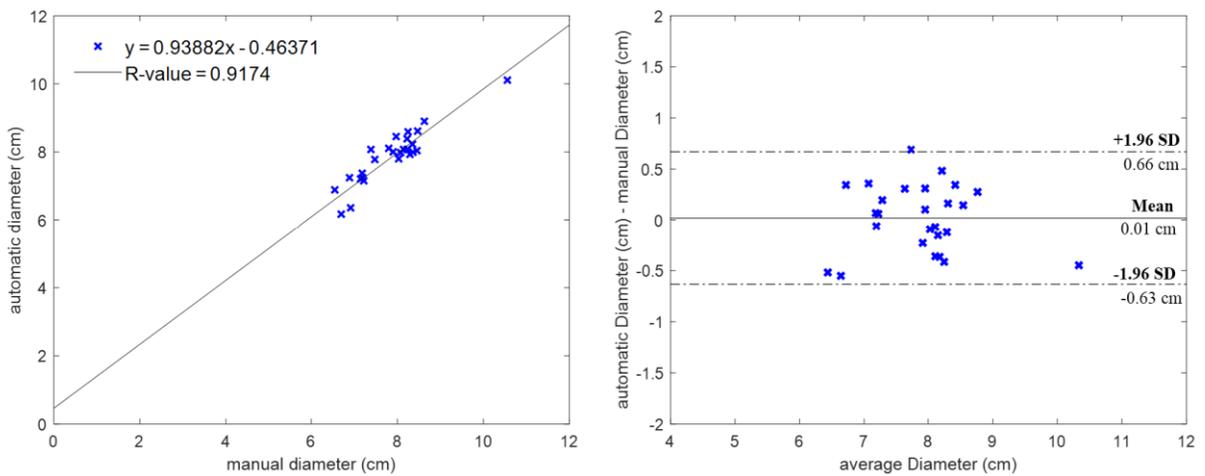

Fig. 8. Correlation graph (left) and Bland-Altman for length in Test set.

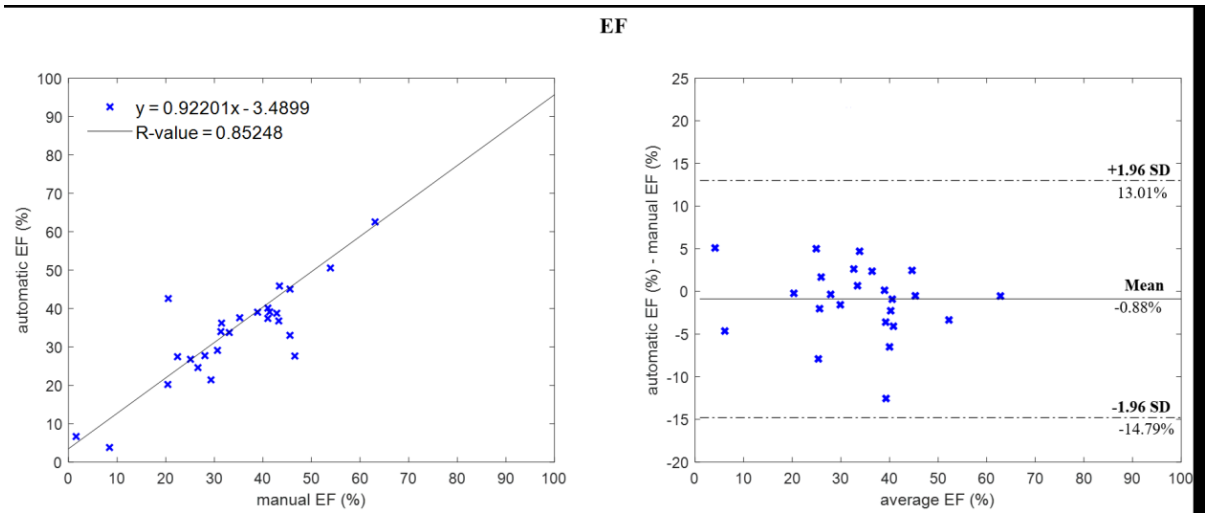

Fig. 9. Correlation graph (left) and Bland-Altman for EF in Test set.

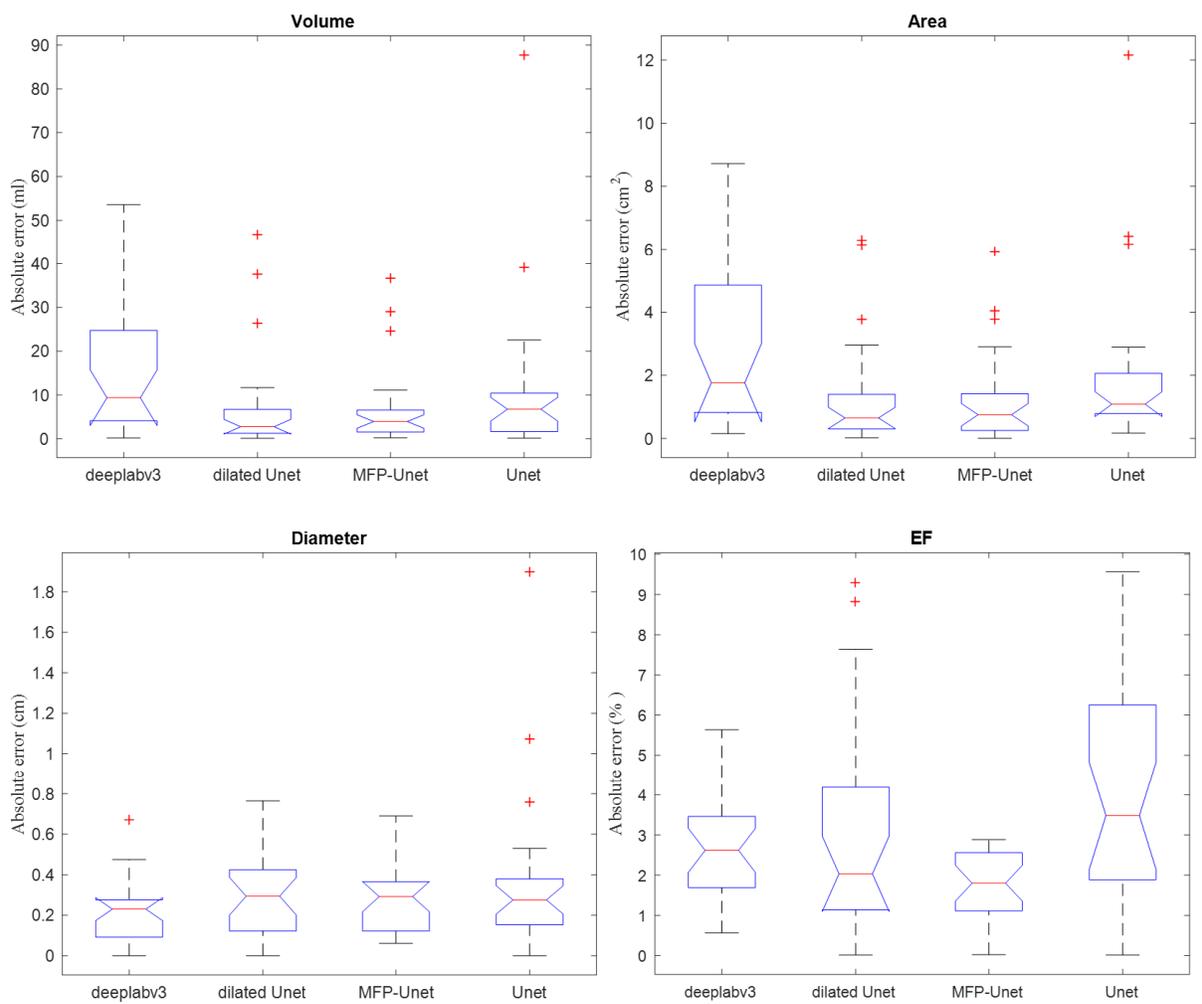

Fig. 10. Box plots of volume, area, length, and EF errors

## DISCUSSION

To summarize this work, this study proposed an effective and efficient hybrid architecture for automatic delineation of the left ventricle from echocardiographic images. A 2D combination of U-net and FPN networks named MFP-Unet, which extracts the LV endocardium in a pixel-to-pixel fashion, was introduced. The proposed network benefits from the robustness of the U-net and flexibility of the FPN that uses the features in all scales for prediction. According to the results, it can be concluded that the proposed approach provides a robust and powerful architecture regarding the capabilities of feature representation in a pyramid.

For instance, a comparison between this approach and other related and state-of-the-art methods demonstrates that this approach generally produces more precise results than U-net, dilated U-net, and deeplabv3. This superiority against the U-net comes from the fact that feature maps with all semantic strengths and all resolutions do not contribute to the segmentation procedure directly in the U-net model. On the other hand, deeplab has a probabilistic module (i.e. conditional random field [49]) which improves the localization of objects. It also segments the instances in the image based on the multi-scale feature maps, as proposed in the method in this study. However, in deeplabv3, the spatial pyramid with various dilation rates are constructed in a pooling layer and the network does not have any copy-concatenation path. While, in the proposed network, the feature pyramids are created by convolution operations and are in the presence of copy-concatenation paths. The results in Fig. 5 also show that the proposed approach has the most agreement with the ground truth segmentation in comparison with other approaches. As it can be seen in Fig. 5, U-net disregards some parts of the LV with more grayscale (like the first and third samples in Fig. 5). Additionally, deeplabv3 misses the LV lateral wall and tends to overshoot LV border (especially on apex) in some cases. Conversely, all three approaches seem to be equally precise in the detection of the septum wall.

An additional layer is concatenated in the input layer of the proposed network which is a preprocessed version of the input image. This study used local thresholding with a threshold value calculated by Niblack's technique which is based on the mean and standard deviation of a window around each pixel. There is also a constant that is multiplied by the standard deviation. This value is obtained in an empirical manner and k=2 shows the best performance in the network.

The proposed network enables significant runtime reduction. For instance, it takes about 1.2 seconds for the classic U-net to segment a test image using the saved weights of a trained model. This value is 1.33 seconds for deeplabv3, according to its high number of parameters. Finally, MFP-Unet needs 0.81 seconds for the same task. It can be predicted that a more efficient implementation of the proposed network will reduce the running time. Even though the classic U-net has fewer convolution layers and thus fewer parameters than MFP-Unet, the lower test runtime of MFP-Unet is not surprising. This is because the convolution kernels of MFP-Unet are dilated kernels which in turn reduce the number of parameters significantly.

The statistical analyses performed on the endocardium segmentation results, i.e. the correlation and Bland-Altman graphs in Figs 6-9 demonstrate that the MFP-Unet results and consequently functional parameters (e.g. EDV, ESV, and EF) have great agreement with manually obtained parameters. Calculated CVs and PRCs show the accuracy and clinical applicability for the evaluation of the LV function. Negligible values of mean and slight values of limits of agreement (±1.96 SD) are illustrative proofs in this regard. Tables 1 and 2 also demonstrate significant reduction in inter-observer variability of volume, area, length, and EF for the proposed model against the manual manner, as the reduction of p-values confirm.

Fig. 10 shows a boxplot representation of errors related to the calculated volume, area, and length by the proposed segmentation method and other related approaches. As it can be seen, the whiskers of the U-net boxplot have the highest distance with the plot. These whiskers can be omitted by training the networks by using a more comprehensive dataset. Additionally, the boxplot of EF, which is the most clinically significant parameter, shows absolute superiority of MFP-Unet against other approaches.

A major concern in the proposed model was the convolutional kernels in the network architecture. There were two choices: regular kernels and dilated kernels. Dilated convolutional kernels, as described in section 2, provides more global receptive field by utilizing a wider field of view. This helps to capture more and more global context from input pixels without increasing the size of the parameters. The classic and dilated U-nets were tested (with dilated kernels at the size of 3) over the prepared dataset, and the dilated U-net achieved more accurate results. Therefore, dilated convolutional kernels are opted to be used in MFP-Unet. The dilation rate was another concern in this context. With a large dilation rate, a lot of small, but significant, details of echocardiographic images that are important in LV segmentation are ignored. Therefore, a low dilation rate, i.e. 2, is used.

The main limitations of the proposed model can be summarized as follows: Although a small training set with an appropriate augmentation can be used to train the MFP-Unet, it is important to have a fairly rich initial training set. It means that the network outperforms when it uses 100 labeled images collected from different patients rather than 100 images from the same patient. In addition, the number of decoding and encoding levels are relatively low, which is due to the low resolution of the input image that limits the number of down-sampling. By having more powerful hardware that provides reasonable training and testing times, input images with higher resolutions can be used, which allow for more numbers of encoding-decoding levels. This in turn increases the number of semantic strengths. Due to the nature of the echocardiographic images, the number of training data cannot be increased more by increasing the augmentation methods. Therefore, increasing the training data is achievable by using more images from more instances.

# CONCLUSION

This study proposed a new supervised learning model for the problem of automatic LV segmentation using 2D echocardiographic images in 4 chamber view. In this paper, the concept of pyramids of features is used in the dilated U-net architecture to create a more precise model. According to the results, the use of features' pyramid and the dilated convolutional kernels show robustness to large and rich training sets. Also, considering all features in all levels of the decoder path provides more accuracy, particularly for challenging slices with low resolution and unclear endocardial border. Within this comprehensive quantitative evaluation, the method showed to be within inter-observer variability, which is an important criterion for its use in a clinical setting. Future works can focus on utilizing the pyramid of feature concept on other deep learning-based segmentation approaches like Densenet and Segnet.

# ACKNOWLEDGMENT

This paper is founded by the Med Fanavaran Plus Co., Karaj, Iran.